\documentclass[twocolumn,apj]{openjournal}
\usepackage[dvipsnames]{xcolor}
\usepackage{amsmath}
\usepackage{makecell}
\usepackage{lmodern}
\usepackage{orcidlink}
\def\xmm{{XMM-{\it Newton }}} 
\def\chandra{{\it Chandra} }  
\def\suzaku{{\it Suzaku} }  
\begin{document}

\title{Mapping the Nearest Ancient Sloshing Cold Front in the Sky with \xmm}

\author{
    Sheng-Chieh Lin\orcidlink{0000-0001-8178-8343}$^{1,2,\ast}$,
    Yuanyuan Su\orcidlink{0000-0002-3886-1258}$^1$,
    Iraj Vaezzadeh\orcidlink{0000-0002-0887-1236}$^3$,
    William Forman\orcidlink{0000-0002-9478-1682}$^4$,
    Elke Roediger\orcidlink{0000-0003-2076-6065}$^5$,
    Charles Romero\orcidlink{0000-0001-5725-0359}$^4$,
    Paul Nulsen\orcidlink{0000-0003-0297-4493}$^4$,
    Scott W. Randall\orcidlink{0000-0002-3984-4337}$^4$,
    John ZuHone\orcidlink{0000-0003-3175-2347}$^4$,
    Ralph Kraft\orcidlink{0000-0002-0765-0511}$^4$ and
    Christine Jones\orcidlink{0000-0003-2206-4243}$^4$
}
\thanks{$^\ast$\href{sli346@uky.edu}{sli346@uky.edu}}
\affiliation{$^1$Department of Physics and Astronomy, University of Kentucky, Lexington, KY 40506, USA}
\affiliation{$^2$Departamento de F\'{i}sica Te\'{o}rica, Universidad Aut\'{o}noma de Madrid, Cantoblanco 28049, Madrid, Spain}
\affiliation{$^3$I. Physikalisches Institut, Universit\"{a}t zu K\"{o}ln, Z\"{u}lpicher Str 77, D-50937 K\"{o}ln, Germany}
\affiliation{$^4$Center for Astrophysics$\vert$Harvard \& Smithsonian, 60 Garden Street, Cambridge, MA 02138, USA}
\affiliation{$^5$E. A. Milne Centre, University of Hull, Cottingham Road, HU6 7RX, Hull, UK}

\begin{abstract}
    The Virgo Cluster is the nearest cool core cluster that features two well-studied sloshing cold fronts at radii of $r \approx 30$ kpc and $r \approx 90$ kpc, respectively. In this work, we present results of \xmm mosaic observations of a third, southwestern, cold front at a radius of $r\approx250$ kpc, originally discovered with {\it Suzaku}. All three cold fronts are likely to be parts of an enormous swirling pattern, rooted in the core.  
    The comparison with a numerical simulation of a binary cluster merger indicates that these cold fronts were produced in the same single event -- likely the infall of M49 from the northwest of Virgo and it is now re-entering the cluster from the south.
    This outermost cold front has probably survived for $2$\,--\,$3$\,Gyr since the disturbance. 
    We identified single sharp edges in the surface brightness profiles of the southern and southwestern sections of the cold front, whereas the western section is better characterized with double edges. 
    This implies that magnetic fields have preserved the leading edge of the cold front, while its western side is beginning to split into two cold fronts likely due to Kelvin-Helmholtz instabilities. 
    The slopes of the 2D power spectrum of the X-ray surface brightness fluctuations, derived for the brighter side of the cold front, are consistent with the expectation from Kolmogorov turbulence.
    Our findings highlight the role of cold fronts in shaping the thermal dynamics of the intracluster medium beyond the cluster core, which has important implications for cluster cosmology. Next-generation X-ray observatories, such as the proposed AXIS mission, will be ideal for identifying and characterizing  ancient cold fronts.
\end{abstract}


\section{Introduction}
\label{sec:intro}

    In the framework of standard $\Lambda$CDM cosmology, mergers and the accretion of substructures are the main channels for the growth of galaxy clusters.
    Such events can significantly impact the dynamical state of a cluster and consequently influence the use of galaxy clusters for cosmology \citep{2022MNRAS.512.3885L, 2024ApJ...977..176L}.
    The merging history of galaxy clusters can be traced by sloshing cold fronts, arising from off-axis or minor mergers \citep{2017ApJ...851...69S}.
    They are characterized by sharp discontinuities in the spatial distribution of surface brightness, temperature, and density of the intracluster medium (ICM), manifested as arcs or spiral patterns in X-ray images \citep{2007PhR...443....1M, 2022hxga.book...93Z}.
    The ``front" forms when low-entropy gas from the cool cores, displaced outward by sloshing, encounters hotter, more diffuse gas at larger radii.
    The interface between the two gas phases creates edges in the surface brightness and discontinuities in thermal properties, with the brighter side being cooler and denser than the fainter side. The characteristic spiral patterns can extend from the core to the outskirts of the galaxy clusters. 

 Cold fronts offer a probe into microscopic plasma physics in intracluster gas \citep{2017ApJ...851...69S}.
    Discontinuities in thermal and metal profiles can inform the effectiveness of transport mechanisms in the ICM \citep{2016JPlPh..82c5301Z}, as diffusion and conduction could broaden the intrinsic width of the front and erase discontinuities. 
    In addition, tangential velocity shear between two gas phases would induce Kelvin-Helmholtz instabilities (KHI), which could develop quickly and distort the cold fronts \citep{2010ApJ...717..908Z, 2011ApJ...743...16Z}.
    Observationally, KHIs often appear as uneven surface brightness edges with ``rolls or horns" \citep{2016MNRAS.457...82S, 2017ApJ...851...69S, 2017MNRAS.468.2506W} and/or multiple edges in the surface brightness profile \citep{2013ApJ...764...60R}. 
    Transport processes and instabilities can be suppressed by magnetic field that is amplified by the sloshing and rearranged to align parallel with the front, also known as ``magnetic draping" \citep{2006MNRAS.373...73L}.
    These microscopic processes are critical for understanding cluster evolution on macroscales, particularly in dissipating and redistributing the kinetic energy from mergers -- an essential factor in utilizing galaxy clusters for cosmology.

\begin{figure*}[t!]
        \centering
        \includegraphics[width=\textwidth]{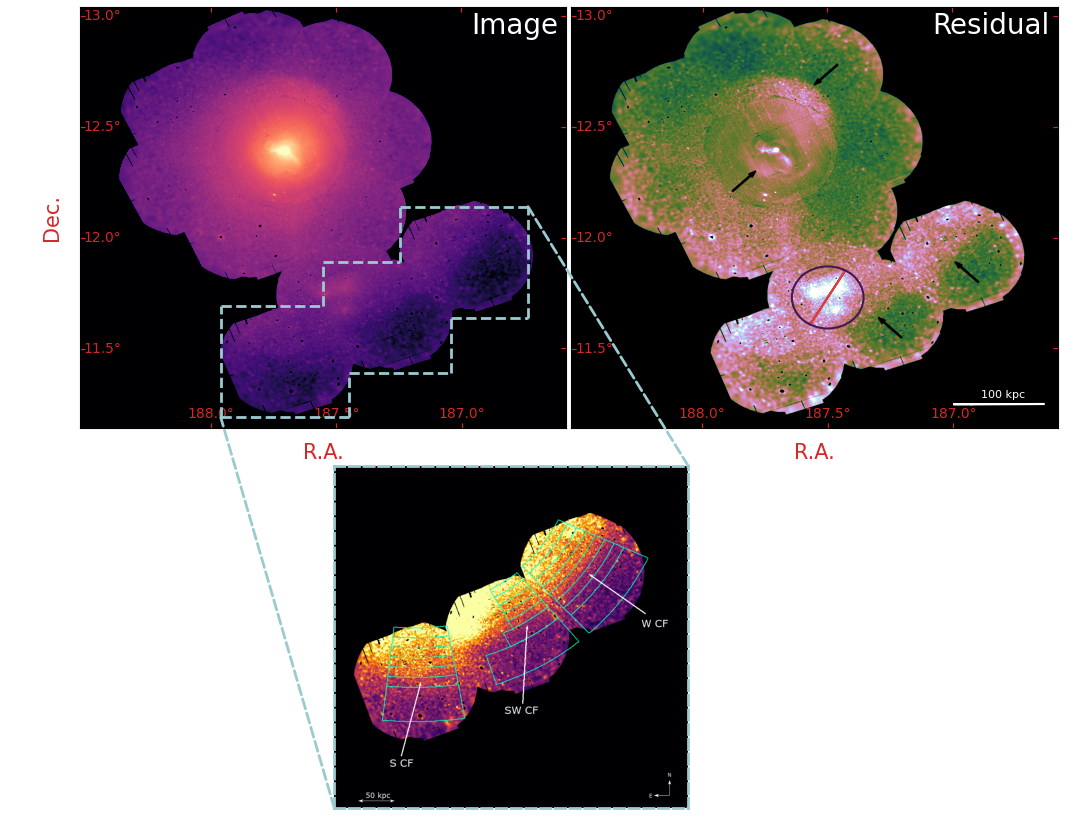}
        \caption{\textit{Left}: Adaptively-smoothed \xmm mosaic image of the Virgo Cluster in the band of $0.5$-$2.0$\,keV after exposure correction and background subtraction. Point sources are removed.
        \textit{Right}: Residual image obtained by dividing the mosaic image by the best-fit beta model.
        The arrows mark the positions of three cold fronts. The two residing in the core region were studied by \cite{2010MNRAS.405...91S}; the outermost cold front is the main focus of this work.
        The region enclosing bright background sources, marked by the black circle, is excluded from the analysis.
        }
        \label{fig:fullimage}
    \end{figure*}

    Nearby cool core clusters such as Virgo and Fornax are ideal targets for studying ICM physics with cold fronts.
    Their proximity allows small structures to be resolved down to a scale that is not achievable with more distant targets.
    The Virgo Cluster harbors two well-studied sloshing cold fronts, residing at radii of $\approx30$ kpc to the southeast and $\approx90$ kpc to the northwest (Fig.\,\ref{fig:fullimage}).
    For the more prominent cold front to the northwest, a very deep \textit{Chandra} observation revealed evident KHI rolls along with suppression of thermal conduction and diffusion \citep{2016MNRAS.455..846W}. 
    The presence of KHI implies a viscosity smaller than the Spitzer value. 
    Similar fine structures have been observed at multiple cold fronts in the Fornax Cluster \citep{2017ApJ...851...69S, 2017ApJ...834...74S}.
    
    Sloshing cold fronts are expected to spread outward over time. However, those at large radii are less well-known due to the lack of archival X-ray observations and the low surface brightness outside cluster centers.   
    The long survival time of such ancient cold fronts could place the most stringent constraints on microphysics. The physical conditions of the ICM outside cluster centers are particularly relevant to cluster cosmology. This region is less influenced by central AGN feedback while remaining shielded from active accretion at the cluster outskirts, making it ideal for measuring cluster masses \citep[e.g.,][]{2004MNRAS.353..457A}.
    The best-studied ancient cold front resides at half the virial radius of the Perseus Cluster, which has remained remarkably sharp for the past 5 Gyr as it moved outward \citep{2012ApJ...757..182S,2018NatAs...2..292W}. \chandra observations reveal that away from the leading edge, the cold front splits into two sharp edges, featuring a hot band between them, likely caused by KHI. The upper limits on the intrinsic widths of both edges are smaller than the Coulomb mean free paths (mfp), suggesting that magnetic draping has prevented the cold front from broadening by diffusion and conduction, whereas KHI starts to disrupt the front \citep{2018NatAs...2..292W}.

    In this paper, we present the results of \xmm observations of the nearest ancient sloshing cold front which lies at an intermediate radius ($250$\,kpc) of the Virgo Cluster.
    This nearest galaxy cluster is modestly massive \citep[$M_{200c}$\footnote{$M_{*}$ indicates the total mass enclosed within a radius of $R_{*}$. Within this radius, the average density is * times the critical density of the Universe at that redshift.}$=1.05\times10^{14}$\,M$_{\odot}$;][]{2017MNRAS.469.1476S} and dynamically young \citep{2007ApJ...655..144M} with a volume-average temperature of kT\,$\approx 2$\,keV.
    At a distance of 16.1\,Mpc \citep[$1$\,arcsec\,$=78$\,pc;][]{2001ApJ...546..681T}, with XMM-{\it Newton}'s spatial resolution of $15$\,arcsec, we can resolve a surface brightness variation down to $1.2$\,kpc. 
    This intermediate-radius cold front was first noted in a previous study using \textit{Suzaku} observations \citep{2017MNRAS.469.1476S}, which mapped the Virgo Cluster in four narrow directions.
    Surface brightness edges are indicated at $r\approx270$\,kpc to the south and at $r\approx230$\,kpc to the west.
    However, the cold front nature cannot be verified with \textit{Suzaku}'s modest angular resolution.
    In this \xmm study, we probe this cold front in more detail using both imaging and spectral analysis. We adopt $R_{\rm vir} \approx R_{\rm 200}=1.05$\,Mpc \citep{2011MNRAS.414.2101U} as the virial radius of the Virgo Cluster.
    The paper is structured as follows. 
    Data preparation and methods are presented in Section \ref{sec:data}.
    The results of the ICM thermal properties are shown in Section \ref{sec:analysis}. 
    The comparison with a binary cluster merger simulation and other findings are discussed in Section \ref{sec:discuss} and summarized in Section \ref{sec:conclude}.
    Throughout this paper, we assume a Solar abundance table of \citet{2003ApJ...591.1220L}, and adopt a cosmology of H$_0$ = $70$ km s$^{-1}$ Mpc$^{-1}$, $\Omega_{\Lambda}$ = $0.73$, and $\Omega_\textrm{m}$ = $0.27$.
    All uncertainties are reported at 1$\sigma$ confidence level.

\section{Observations and Data Reduction}
\label{sec:data}

    The intermediate-radius cold front in Virgo has been mapped with \xmm with three pointings.
    The mosaic observations cover the southern and western edges of the cold front, which was previously detected with \textit{Suzaku}. In addition, assuming these two edges are part of the same cold front extending from the south to the west, we targeted a southwestern region bridging the two \suzaku fields. 
    A recent eROSITA study, mapping the entire Virgo Cluster, has confirmed this assumption \citep[also see Fig.\,\ref{fig:simulation}-right]{2024A&A...689A.113M}.
    The \xmm observations were made in July 2020. 
    Details of the observations are listed in Table \ref{tab:summary}.

    \begin{table*}
        \centering
        \caption{Summary of \xmm pointings of the intermediate-radius cold front}

\begin{tabular}{cccccccc}
        
            Obs. ID & Date & Total Exposure (ks) & Net MOS/pn (ks)  & R.A. & Dec. & PI & Label \\
            \hline\hline 
            0863810101 & 2020-07-08 & 28 & 26/21 & 12 29 08.12 & 11 39 13.6 & Su & S \\
            0863810201 & 2020-07-08 & 34 & 32/26 & 12 27 50.62 & 11 54 57.6 & Su & SW \\
            0863810301 & 2020-07-05 & 44 & 33/27 & 12 30 42.66 & 11 27 17.1  & Su & W \\
        \end{tabular}
  \label{tab:summary}
    \end{table*}

    The \xmm EPIC data were reduced using the Science Analysis System (SAS) packages version 20.0.0 together with the \xmm Extended Source Analysis Software package \citep[XMM-ESAS,][]{2004ApJ...610.1182S}.
    MOS data and pn data were reprocessed and calibrated using \texttt{\textbf{emchain}} and \texttt{\textbf{epchain}}.
    \texttt{\textbf{mos-filter}} and \texttt{\textbf{pn-filter}} were used to remove flares.
    The event files were filtered by setting FLAG $= 0$ and PATTERNS $<= 12$ for MOS data and FLAG $= 0$ and PATTERNS $<= 4$ for pn data.
    The out-of-time (OOT) events were removed from both the images and spectra for pn data.
    The total cleaned time was approximately $75$\,ks. The cleaned exposure times of individual observations are listed in Table \ref{tab:summary}.
    Point sources detected by the SAS task \texttt{\textbf{cheese}} were excised with a count rate threshold of $5\times10^{-15}$\,ergs\,cm$^{-2}$\,s$^{-1}$ and a minimum source separation of $20''$.

    We generated images in the energy band of $0.5$\,--\,$2.0$\,keV.
    The particle background images were created using \texttt{\textbf{mos\_back}} and \texttt{\textbf{pn\_back}}.
    \texttt{\textbf{comb}} was used to create background-subtracted, vignetting-corrected images with point sources excised.
    Finally, \texttt{\textbf{adapt}} was used to mosaic these images with an adaptive smoothing binning of $2$ and a minimum count of $50$ per bin.
    In addition to our newly obtained \xmm observations, we included archival observations of the inner region of the Virgo center as presented in \cite{2010MNRAS.405...91S}. 
    A final mosaic image of all the existing \xmm observations of this field is shown in Fig.\,\ref{fig:fullimage}.
    In the southwestern pointing, three bright background sources visibly overlap with the Virgo ICM in projection.
    Two of them are background clusters at $z=0.085$ \citep{2021A&A...650A.104C} and $z=0.069$ \citep{2009ApJS..183..197W}, respectively, and the bright source to the south is an active galactic nucleus. These sources were excluded in the analysis. 

    We considered two sources of background: the astrophysical background (AXB) and the non-X-ray background (NXB). The filter wheel closed (FWC) data were utilized to constrain the NXB as detailed in Appendix\,\ref{sec:bkg}. The tasks \texttt{\textbf{rmfgen}} and \texttt{\textbf{arfgen}} were used to generate redistribution matrix files (RMFs) and auxiliary response files (ARFs), respectively, for spectra analysis.

\section{Analysis and Results}
\label{sec:analysis}

    \subsection{Image Analysis}
    \label{subsec:imganalysis}
        In Fig.\,\ref{fig:fullimage}, the background-subtracted exposure-corrected mosaic image depicts surface brightness edges at a radius of $\approx 250$\,kpc spanning roughly $80^{\circ}$ from the west to the south.
        To further highlight this outer cold front, we produced a residual image by dividing the mosaic image by the best-fit beta model defined as,
        \begin{equation}
            \label{eq:sbmod}
            S(r) = (1 + (r/R_0)^2)^{-3\beta+0.5},
        \end{equation}
        where $r$ is the projected distance from the cluster center.

        To quantify the surface brightness (SB) distribution, we extracted SB profiles from regions in the west (W), south-west (SW), and south (S) directions, as shown in Fig.\,\ref{fig:thermalprofile}. 
        The radial profiles were extracted in annular elliptical sectors centered 
        on M87 and chosen to best match the curvature of the cold front.
        Both the image and spectral analysis are limited to the innermost $10'-12'$ of the \xmm fields.
        The extraction was conducted using the Python package \textbf{pyproffit} \citep{2020OJAp....3E..12E}. 
        We fitted the surface brightness profiles with a 3D broken power law with a density jump formulated as,
        \begin{equation}
            \label{eq:singleedge}
            n_e = 
            \begin{cases}
                n_{e0} \left( \dfrac{r}{r_c} \right)^{-\alpha_{1}} & r < r_c\\
                \dfrac{n_{e0}}{C} \left( \dfrac{r}{r_c} \right)^{-\alpha_{2}} & r \ge r_c\\
            \end{cases},
        \end{equation}
        where $r$ is the elliptical radius, $\alpha_{i}$ is the power-law index, $r_c$ is the corresponding radius of the edge, $C$ denotes the density jump, and $n_{e0}$ is the normalization.
        The 3D density model was projected along the line-of-sight assuming elliptical symmetry \citep[see appendix in][]{2009ApJ...704.1349O}.
        The best-fit parameters were estimated using the maximum likelihood method with the \textbf{iminuit} package \citep{iminuit}, and the uncertainties were obtained using the \texttt{\textbf{MINOS}} function in \textbf{iminuit}.
        The best-fit results are shown in the upper panel of Fig.\,\ref{fig:thermalprofile}, and the best-fit parameters are listed in Table~\ref{tab:singleparam}.
        The edge radii in the W and SW regions are in the range of $50'$\,--\,$52'$, while in the S region, the edge is more extended outward to $55'$.

        \begin{table*}
            \centering
            \caption{Best-fit broken power-law model parameters. For the W wedge, we also show the double-edge parameters. The edge radii are the projected distances from M87.
            }
            \begin{tabular}{cccc}
                Obs. & $r_c$\,(arcmin) & $C$ & $\chi^2/dof$\\
                \hline
                S & $55.60^{+0.08}_{-0.01}$ & $1.22^{+0.05}_{-0.05}$ & $46.18/40$ \\
                SW & $49.98^{+0.36}_{-0.19}$ & $1.28^{+0.09}_{-0.09}$ & $53.56/43$ \\
                W & $52.45^{+0.14}_{-0.15}$ & $1.39^{+0.06}_{-0.06}$ & $88.59/46$ \\
                W (double) & $52.36^{+0.14}_{-0.04}$, $54.78^{+0.01}_{-0.01}$ & $1.43^{+0.08}_{-0.05}$, $1.51^{+0.14}_{-0.14}$ & $69.45/43$\\
                \hline\hline
            \end{tabular}
            \label{tab:singleparam}
        \end{table*}

        Sloshing cold fronts can display multiple proximate edges following the establishment of KHIs \citep{2013ApJ...764...60R}.
        A similar effect, as seen in Perseus, is the split of a cold front into two edges by KHI. 
        To identify potential multi-edge features, we fitted a 3D double-edge broken power-law \citep{2017MNRAS.467.3662I} to the SB profiles in all three wedges,
        \begin{equation}
            \label{eq:doubleedge}
            n_e = 
            \begin{cases}
                n_{e0} \left( \dfrac{r}{r_{c1}} \right)^{-\alpha_{1}} & r < r_{c1}\\
                \dfrac{n_{e0}}{C_1} \left( \dfrac{r}{r_{c1}} \right)^{-\alpha_{2}} & r_{c1} < r \le r_{c2}\\
                \dfrac{n_{e0}}{C_1 C_2} \left( \dfrac{r_{c2}}{r_{c1}} \right)^{-\alpha_{2}} \left( \dfrac{r}{r_{c2}} \right)^{-\alpha_{3}} & r_{c2} \le r \\
            \end{cases}.
        \end{equation}
        The symbols follow the same naming convention as in Eq.~\ref{eq:singleedge}.

        To confirm the detection of double edges, we compared both SB models using the Bayesian information criterion \citep[BIC;][]{1978AnSta...6..461S} defined as,
        \begin{equation}
            BIC = \chi^2 + k \ln{n},
        \end{equation}
        where $k$ is the number of free parameters and $n$ is the number of data points.
        In general, a model with a smaller BIC value is preferred.
        For our model selection, we chose the double-edge model if $\Delta BIC = BIC_{\text{single}} - BIC_{\text{double}} > 6$ \citep{2010A&A...511A..85P}.
        In the S and SW wedges, we found no significant differences between the BICs.
        In the W wedge, the difference $\Delta BIC = 7.05$ is statistically large such that the double-edge model is preferred over the single-edge model.
        For the W wedge, the best-fit parameters of the double-edge model are shown in the bottom row of Table~\ref{tab:singleparam}.

    \subsection{Spectral Analysis}
        The spectra were extracted from sectors of elliptical annuli aligned with the cold front as shown in Fig.~\ref{fig:fullimage}.
        We performed spectral analysis for all three wedges across the cold front using the XSPEC package \citep{1996ASPC..101...17A,2001ASPC..238..415D}.
        Each spectrum was binned to minimum of $25$ counts per bin.
        The spectral fitting was restricted to $0.5$\,--\,$10.0$\,keV, and the chi-squared statistic was used during fitting.
        We also conducted the fitting with the binning of a minimum of $1$ count per bin and with C-statistics, which yield consistent results.
       
        The thermal properties of the ICM in each region were modeled by assuming an absorbed single-temperature collisional plasma, $\texttt{phabs}\times\texttt{apec}$, with its redshift fixed at $z=0.00428$, the redshift of the central galaxy M87.
        The metallicity was fixed at $0.3\,Z_{\odot}$.
         Galactic absorption is modeled using \texttt{phabs} with the column density of hydrogen fixed at $n_H = 1.3 \times 10^{20}$ cm$^{-2}$ \citep{2005A&A...440..775K}.

\begin{figure*}
            \centering
            \includegraphics[width=0.85\textwidth]{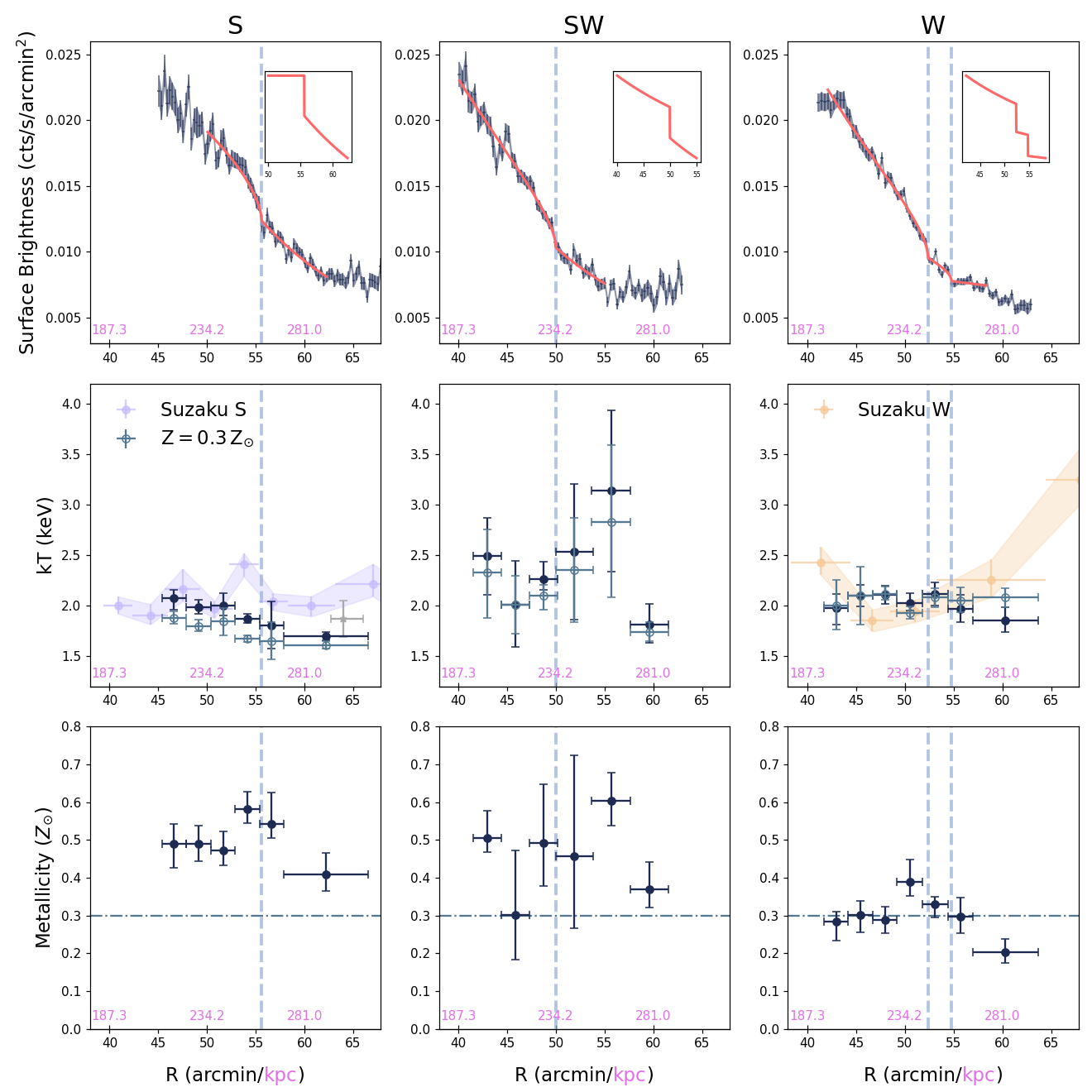}
            \caption{
            Surface brightness, projected temperature and projected metallicity profiles along the S, SW, and W wedges.  
            The radii of cold fronts are denoted by the dashed lines (W wedge exhibits two edges).
            The x-axis labels the projected distance R from M87 in both arcmin (black) and kpc (purple).
            \textit{Upper}: surface brightness profiles together with the best-fit models in red.
            The inset shows the best-fit models of 3D density profiles.
            \textit{Middle}: projected temperature profiles overlaid with the measurements made by \suzaku \citep{2017MNRAS.469.1476S}. The error bars include both statistical and systematic uncertainties.
            The filled points show the best-fit temperatures when varying the metallicity in the \texttt{apec} model, and the open points are the results with metallicity fixed at $0.3$.
            The gray point at $R \approx 65'$ shows the result from an extraction region matching the \suzaku observation area at the same radius.
            \textit{Bottom}: best-fit metallicity profiles when varying the metallicity in the \texttt{apec} model. The horizontal dashed-dotted line marks Z$=0.3\,$Z$_{\odot}$, corresponding to the open points in the middle panel.
            }
            \label{fig:thermalprofile}
        \end{figure*}

        We adopted the same model as for the background analysis (Appendix\,\ref{sec:bkg}) including the NXB, AXB, and ICM components.
        The resulting temperature profiles are shown in Fig.\,\ref{fig:thermalprofile}. The error bars include both statistical and systematic uncertainties (Appendix\,\ref{sec:sys}). 
        With the projected temperature profile, we observe a temperature jump across the cold front for the SW direction, from $2.2$\,keV to $2.5$\,keV, while those of the S and W directions vary mildly or are consistent within the uncertainties. We notice discrepancies in the projected temperature measurements between \xmm and {\it Suzaku}, which may be attributed to the different extraction regions used in the analyses and/or the systematic difference between the two telescopes. 
       When extracting spectra from regions matching the \suzaku field, we found more consistent measurements (gray point in the S panel of \ref{fig:thermalprofile}).
       
\begin{figure}
            \centering
            \includegraphics[width=\columnwidth]{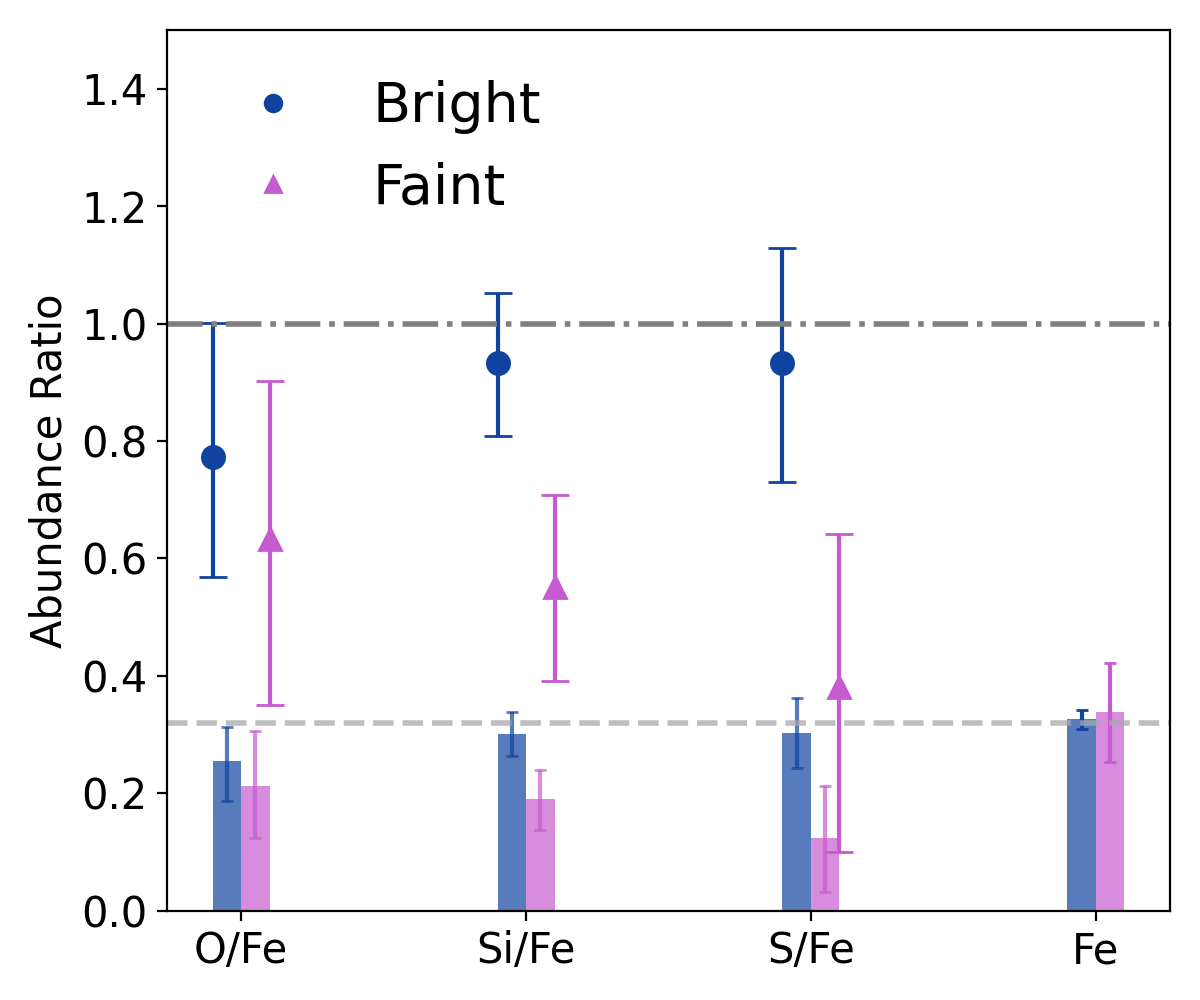}
            \caption{The metal abundance and the abundance ratios of O, Si, S, with respect to Fe. The measurements are made for the bright side (blue) and the faint side (magenta). 
            The circles show the abundance ratios, and the bars show the abundances.
            The error bars include statistical and systematic uncertainties (Appendix\,\ref{sec:sys}). The dashed-dotted line marks the ratio of $1$. The dashed line marks the Fe abundance of $Z\approx 0.32$.
            }
            \label{fig:abund}
        \end{figure}

        We were unable to constrain the metallicity of each annular sector individually.
        Instead, we measured the chemical composition of the entire region on each side of the cold front, using \texttt{vapec}, a variant of \texttt{apec}, initially allowing the abundances of 13 chemical elements free to vary. 
        We found it necessary to link most $\alpha$ elements to O, and Ni to Fe. 
        The estimated abundances of different elements for both bright and faint sides are shown in Fig.~\ref{fig:abund}.

    \subsection{\textit{Deprojections} of Thermal Profiles}

\begin{figure*}
            \centering
            \includegraphics[width=0.85\textwidth]{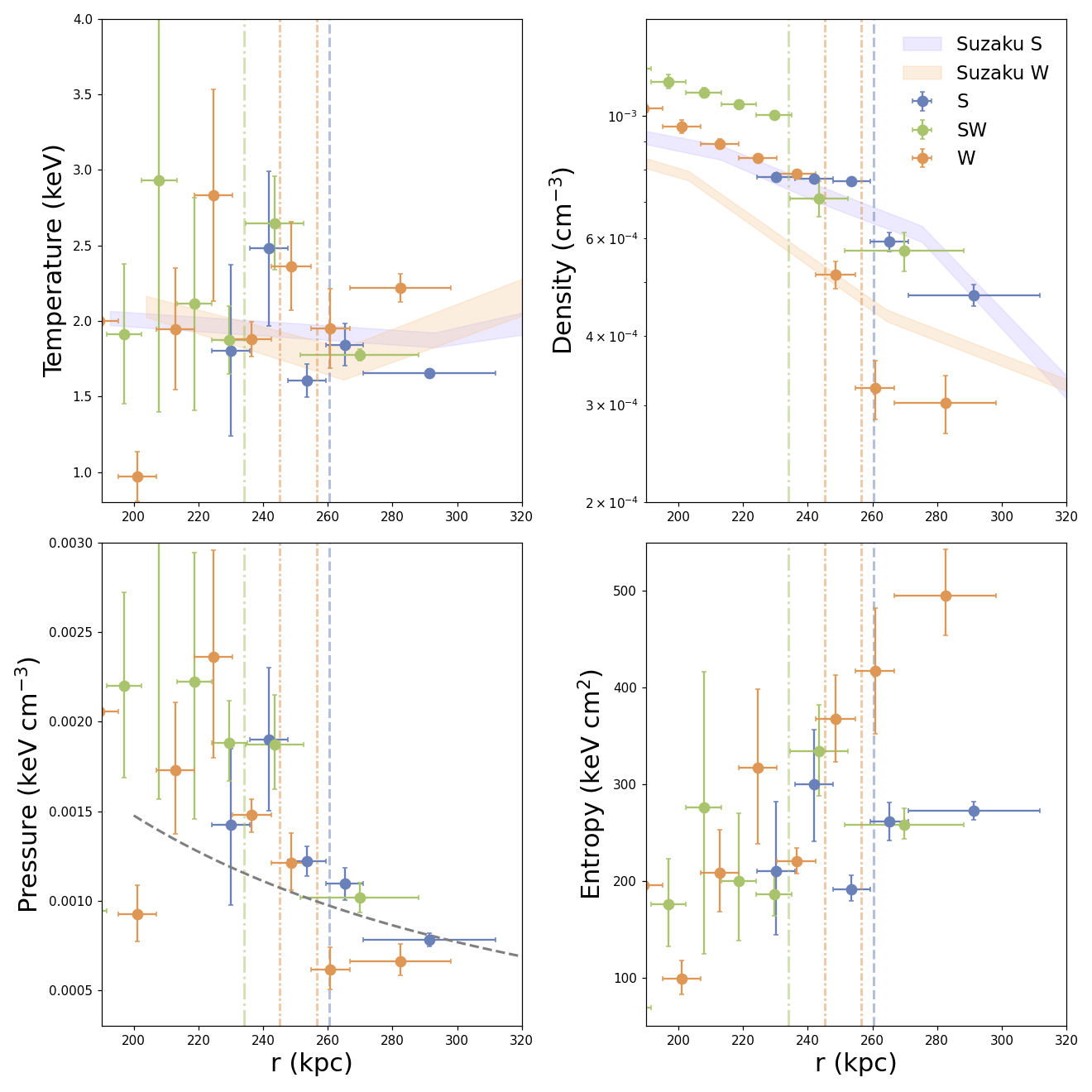}
            \caption{
                \textit{Deprojected} profiles of temperature, density, pressure, and entropy along the S (blue), SW (green), and W (orange) directions. Shaded areas are measurements made by \textit{Suzaku} \citep{2017MNRAS.469.1476S}.
                Dashed vertical lines mark the positions of SB edges in three directions.
                In the lower left panel, we include the empirical pressure profile given by Eq.~1 in \cite{2017MNRAS.469.1476S}.
            }
            \label{fig:deprojth}
        \end{figure*}

        We estimated the \textit{deprojected} temperature profiles using the extracted spectra from the regions as shown in the zoom-in plot in Fig.~\ref{fig:fullimage}.
        We followed the deprojection procedure described in \cite{2007MNRAS.381.1381S}, where deprojection is performed upon the spectra, incorporating background spectra.
        The background spectra were simulated assuming the best-fit NXB and AXB models from the previous projected analysis using XSPEC's \texttt{\textbf{fakeit}}.
        As shown in Fig.\,\ref{fig:deprojth}, the \textit{deprojected} temperature profiles show more pronounced discontinuities across the cold front in all three directions (see \S\ref{sec:discuss} for details).
        The density estimates given by the deprojection method  
        can be subject to several systematic uncertainties, including the assumption of true volume and the projection effect from the gas at the outer bins.
        We thus derived the density profiles directly from the surface brightness profiles.
        The SB profiles offer an estimate of the shape of the density profile.
        To determine the absolute values, we utilize the best-fit normalization of the \texttt{apec} model in the projected spectra analysis, which is the emission measure (EM) scaled with the cosmological distance.
        Assuming the \textit{deprojected} temperature and density are uniformly distributed in each annular bin, the EM can be reduced to the integration of density squared along the line-of-sight (LOS)
        \begin{equation}
            \text{norm}_{\texttt{apec}} \propto EM \sim A \int_{\text{LOS}} n_e n_H dl,
        \end{equation}
        where $A$ is the area of a region, $n_e$ is the electron density, $n_H(=n_e/1.2)$ is the hydrogen density. 
        With norm$_{\texttt{apec}}$, we are able to estimate the density profiles with Eq.~\ref{eq:singleedge} and Eq.~\ref{eq:doubleedge}. With \textit{deprojected} temperature and density profiles, we further derive the pressure ($P=nkT$) and entropy ($s=kT / n^{2/3}$) profiles as shown in Fig.\,\ref{fig:deprojth}.

\section{Discussion}
\label{sec:discuss}

        \begin{figure*}
        \centering
        \includegraphics[width=\textwidth]{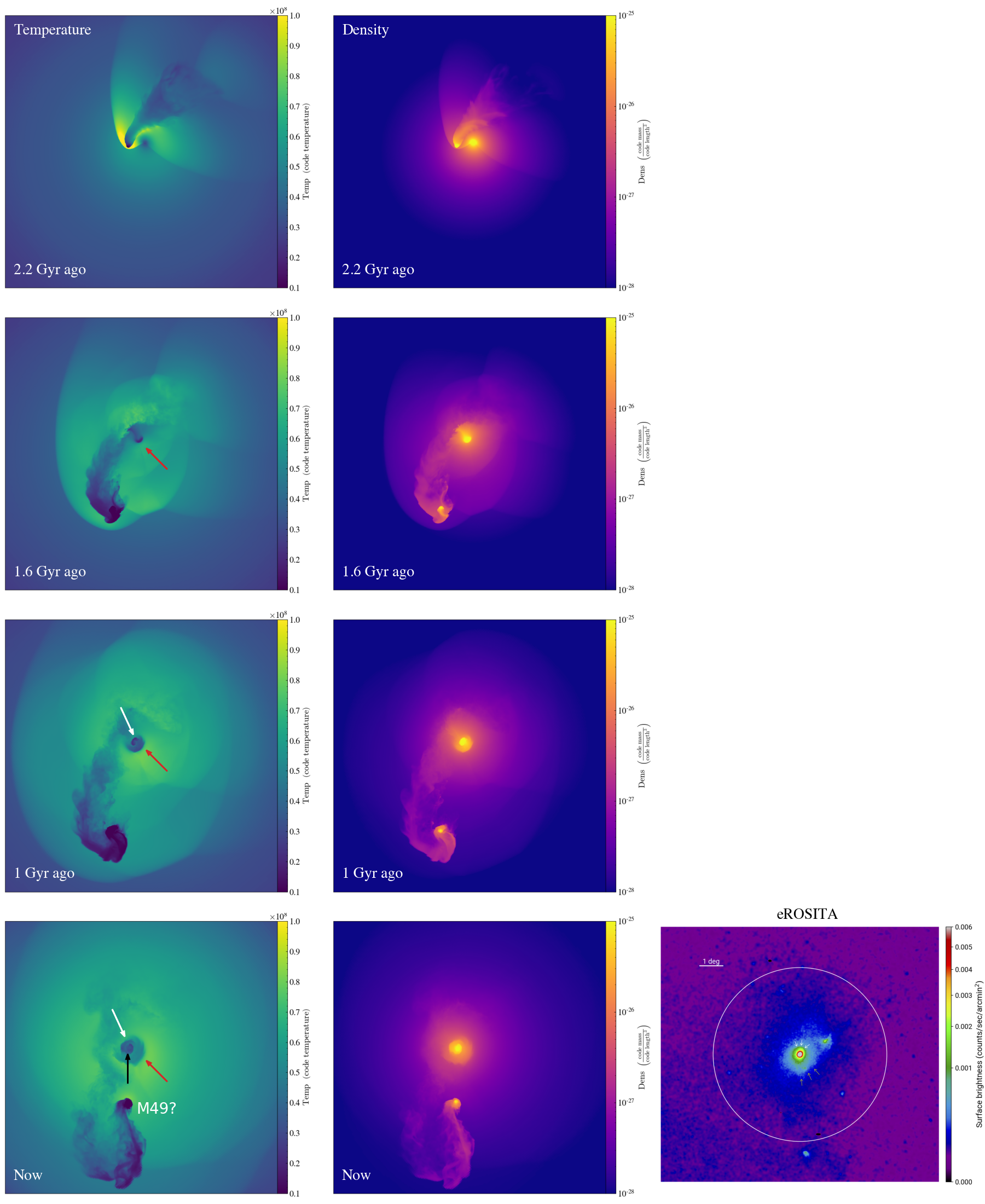}
 \caption{The merging history inferred for the Virgo Cluster. \textit{Left}: Snapshots of temperature maps of a binary cluster merger, derived from a numerical simulation  \citep{2022MNRAS.514..518V}, featuring the development of sloshing cold fronts. \textit{Middle}: Same as the left column but for the density map. Both temperature and density maps are X-ray emission weighted. The 4 panels from top to bottom: the first core passage of an
infalling substructure occurs 2.2 Gyr ago; the first sloshing cold front (red arrow) appears 1.6 Gyr ago as the substructure is approaching
apoapsis; the second sloshing cold front (white arrow) develops 1 Gyr ago when the substructure is turning around; all three sloshing cold
fronts (red, white, and black arrows) are visible as the substructure, likely to be M49, is re-entering the virial radius of the main cluster, which provide the best match to the X-ray image of the Virgo Cluster. \textit{Right}: Point source excised, adaptively smoothed eROSITA X-ray image of the Virgo Cluster. The white circle marks the r$_{200c}$ radius. M49 is visible to the south of the Virgo Cluster. This image is taken from \cite{2024A&A...689A.113M} with permission.}
    \label{fig:simulation}
    \end{figure*}

    \subsection{An ancient sloshing cold front in Virgo}

        The mosaic image of \xmm reveals three sloshing cold fronts in Virgo at radii of $\approx30$\,kpc to southeast, $\approx90$\,kpc to northwest, and $\approx250$\,kpc to southwest.
        The outermost one is the largest known sloshing cold front in Virgo.
        It spans from south to west with best-fit edges at radii of $270$\,kpc to the south (S), $243$\,kpc to the south-west (SW), and $254$\,kpc to the west (W). 
        It is likely an ancient sloshing cold front, induced by an off-axis merging event early on, and has been propagating (or rolling) outward ever since.
        To gain more insight into its merging history, we make use of a hydrodynamic simulation of a binary merger presented in \cite{2022MNRAS.514..518V}.

        Snapshots of several key moments are highlighted in Fig.\,\ref{fig:simulation}. 
        The first pericenter passage occurs at $1.5$\,Gyr after the simulation started (``$2.2$\,Gyr ago" in Fig.\,\ref{fig:simulation}) and the distance between the cores of the main cluster and the subcluster is approximately $200$\,kpc.
        The passage of the substructure separates the peaks of dark matter and ICM of the main cluster since the intracluster gas is subject to the ram pressure of ambient gas while dark matter is not.
        As the subcluster passes through and the direction of the gravitational pull changed, the intracluster gas no longer feels the ram pressure, beginning to fall back to the gravitational potential minimum, overshooting, and continuously propagating outward.
        The first cold front starts to emerge at $\approx1.9$\,Gyr and becomes prominent at $\approx2.1$\,Gyr (``1.6\,Gyr ago" in Fig.\,\ref{fig:simulation}). 
        Proceeding from there, the cold gas continues swirling to larger radii, and the spiraling patterns are caused by the angular moment transferred from the substructure. The second cold front appears at $\approx2.8$\,Gyr (``1\,Gyr ago" in Fig.\,\ref{fig:simulation}) as the subcluster starts to turn around.
        Comparing to the residual map (Fig.\,\ref{fig:fullimage}-right), we found the best match is the snapshot at $3.7$\,Gyr ($2.2$\,Gyr after the first pericenter passage, ``now" in Fig.\,\ref{fig:simulation}).
        The three cold fronts at different radii, facing the expected directions, are clearly visible in this snapshot. The infalling subcluster is re-entering the virial radius of the main cluster from the south, which displays a prominent merging cold front facing north and a stripped tail pointing to the southwest.
        
        The X-ray morphology of the Virgo ICM may have been shaped by the interactions between the main cluster and a number of substructures during a complicated merging history. However, in a simplified binary merger scenario, the most likely candidate responsible for these sloshing cold fronts is M49, which resides south of M87 and near the virial radius. It shows a prominent merging cold front facing north \citep{2019AJ....158....6S} and a stripped tail pointing to the southwest \citep{2024A&A...689A.113M,2024A&A...690A.195S}. We stress that this is not a simulation specifically tailored to the Virgo Cluster. In the simulation setup, the main cluster and the subcluster have masses of $5\times10^{14}$\,M$_{\odot}$ and $5\times10^{13}$\,M$_{\odot}$, respectively, whereas the mass of Virgo is $1.05\times10^{14}$\,M$_{\odot}$ \citep{2017MNRAS.469.1476S} and that of M49 is $4.6\times10^{13}$\,M$_{\odot}$ \citep{2019AJ....158....6S}.
        The comparison with the simulation can only be taken as a qualitative probe of the merging history of Virgo and the formation of its sloshing cold fronts.

        In observations, the sloshing timescale can be estimated assuming a buoyant restoring force applied to the cool, dense gas from the cluster core.
        The corresponding buoyancy (Brunt-V\"{a}is\"{a}l\"{a}) frequency is given by,
        \begin{equation}
            N = \Omega_{K} \sqrt{\dfrac{1}{\gamma}\dfrac{d\ln{s}}{d\ln{r}}},
        \end{equation}
        where $\Omega_{K}=\sqrt{GM/r^3}$ is the Keplerian frequency, $s$ is the gas entropy, and $\gamma=5/3$ is the adiabatic index.
        Although the Brunt-V\"{a}is\"{a}l\"{a} frequency specifically describes the vertical oscillation of fluid elements in a stratified medium, it can serve as a reasonable approximation for the sloshing timescale \citep{2003ApJ...590..225C,2017ApJ...851...69S}.
        Using the best-fit entropy profile of Virgo from \textit{Suzaku} observations \citep[][Eq.~2]{2017MNRAS.469.1476S}, the sloshing period is estimated to be $3.3$\,Gyr old, which is greater than that inferred from numerical simulation ($2.2$\,Gyr).

\subsection{The split in the ancient cold front}

The observations reveal that the surface brightness profiles across this outer cold front, to S and SW, can be well fitted by a single-edge (broken powerlaw density) model, while that toward the W can be better fitted with a double-edge model (Fig.\,\ref{fig:thermalprofile}). The temperature profile across the W front implies an elevated temperature between the two edges (Fig.\,\ref{fig:deprojth}). We speculate that this sloshing cold front is beginning to split into two cold fronts from SW to W by KHI, resembling the ancient cold front in Perseus \citep{2018NatAs...2..292W}.
   The binary merger simulation from \cite{2022MNRAS.514..518V} that we used is for an inviscid unmagnetized ICM, in which KHI can quickly develop. 
   KHI features are visible in the snapshot that best match the current X-ray image of Virgo -- the portion of the outer sloshing cold front, away from the leading edge, starts to split into two cold fronts with an elevated temperature appearing between the two cold fronts, in agreement with our observation. 

    It is worth noting that the ancient cold front in Perseus lies at $700$\,kpc from the cluster center, almost half the virial radius; the two edges are separated by nearly $100$\,kpc. In contrast, the cold front in Virgo resides at $250$\,kpc ($0.25$\,R$_{\rm vir}$) and the two edges are separated by $\approx 10$\,kpc. This Virgo cold front is likely a younger version of the cold front in Perseus, allowing us to witness the early-phase of the splitting. This is consistent with their merger histories inferred from numerical simulations that the front in Perseus is $\approx5$\,Gyr old \citep{2018NatAs...2..292W}, while that in Virgo is $2$\,--\,$3$\,Gyr old, and the fact that Virgo is a dynamically younger cluster. With those two ancient cold fronts and the two inner cold fronts in Virgo, we witness the evolution of a sloshing cold front -- it was formed and subsequently disrupted as it propagated outward during the cluster' relaxation.

    \subsection{Thermal and chemical properties across cold front}

        The \textit{deprojected} temperature profiles in all three directions show pronounced temperature jumps at the edges.
        For S, SW and W, the corresponding temperature jumps, $kT_{\rm faint}/kT_{\rm bright}$, are $1.15\pm0.12$, $1.42\pm0.24$, and $1.25\pm0.17$, consistent with the level of density drop (Fig.\,\ref{fig:deprojth}).
        The derived pressure profiles are thus continuous across the edge.
        The derived entropy profiles appear to be discontinuous, as the gas inside the cold front has a lower entropy than the gas outside the front.
        Overall, the thermal distribution is consistent with the expectations for cold fronts.

         At radii larger than the faint side of the cold front, we measure a sudden drop in temperature in all three directions (Fig.\,\ref{fig:deprojth}),    
        accompanied by the decline of the ICM density at larger radii. The  pressure drops abruptly at $\gtrsim 260$\,kpc,
     with a pressure
        gradient $\partial P/\partial r$ more pronounced than
        that derived from the empirical global pressure profile of the Virgo Cluster \citep[Eq.~1]{2017MNRAS.469.1476S} at similar radii.
       The thermal pressure discontinuity may imply the presence of non-thermal pressure support at this radius.
        If the non-thermal pressure is mainly due to the magnetic field, its magnitude can be estimated through,
        \begin{equation}
            B = \sqrt{8 \pi \Delta P},
        \end{equation}
        from which we derive field strengths of $2.00\pm0.50$\,$\mu$G, $3.54\pm 0.77$\,$\mu$G and $3.24\pm 0.65$\,$\mu$G for the S, SW, and W, respectively.
        On the other hand, if the turbulence is the main reason for the non-thermal pressure, the corresponding 3D velocity dispersion would be
        \begin{equation}
            \delta_v = \sqrt{\dfrac{3 P_{tur}}{\rho}},
        \end{equation}
        where $\rho = \mu m_p n$ is the gas density calculated from the estimated \textit{deprojected} density n, the proton mass $m_p$ , and the mean molecular mass per particle assumed at $\mu=0.59$.
        We obtain $319.86\pm80.48$, $516.98\pm114.35$, and $630.74\pm131.03$\,km s$^{-1}$ for the S, SW, and W, respectively.
        Compared to the sound speeds at that radius, the corresponding Mach numbers $\mathcal{M}$ in the three directions are $0.48$ (S), $0.75$ (SW), and $0.87$ (W).
        However, we stress that the deprojected pressure profiles derived here are subject to significant systematic uncertainties and should be interpreted with caution.

        We measure various chemical elements across this cold front far from the cluster center.
        The metal abundance (O, Si, S, and Fe), and the abundance ratios with respect to Fe on the fainter and brighter sides are shown in Fig.\,\ref{fig:abund}.
        The Fe abundances on both sides are consistent with 0.3\,Z$_{\odot}$.
        All the constrained abundance ratios (X/Fe) appear to be sub Solar. 
        ICM parameters measured on the fainter side represent the properties at the radius of the cold front, while those on the brighter side may represent gas at an inner region of the cluster where the sloshing cold front gas originates. The comparison between the brighter and fainter sides of the cold front, in terms of Fe and abundance ratios, may reflect the level of uniformity in the chemical composition over a large span of radii in the ICM. In this regard, our result supports a well-mixed ICM metallicity \citep{2023MNRAS.526.6052S,2022MNRAS.516.3068S}, which may indicate that chemical elements were expelled, through stellar and AGN feedback, into the intergalactic medium during the early epoch of cluster formation. 

    \subsection{Intrinsic width of the cold front width}

        Comparing the intrinsic width of the cold front with the Coulomb mean free path (mfp), one can determine the effectiveness of the thermal diffusion. If the width is several times the mfp, 
        the rate of diffusion is consistent with being controlled by Coulomb collisions.
        Otherwise, the diffusion may be suppressed by the oriented magnetic field along the cold front's interface, resulting in a narrower width.

        Mathematically, the electron mfp is calculated via,
        \begin{equation}
            \lambda = \dfrac{3^{3/2} (kT_{e})^2}{4 \sqrt{\pi} n_{e} e^4 \ln{\Lambda}},
        \end{equation}
        where $kT_{e}$ is the electron temperature, $n_e$ is the electron number density, and $e$ is the electron charge.
        The Coulomb logarithm $\ln{\Lambda}$ is computed as follows,
        \begin{equation}
            \ln{\Lambda} = 35.7 + \ln{\left(kT_{e}/\sqrt{{n}_{e}}\right)}.
        \end{equation}
        Adopting the \textit{deprojected} temperature and density values at the edge of the W, the mfp inside and outside the cold front are $\lambda_{i}=1.88$\,kpc and $\lambda_{o}=2.36$\,kpc, respectively.
        To assess the degree of diffusion from the denser to the more diffuse side, we calculate the mfp crossing the edge using
        \begin{equation}
            \lambda_{i\rightarrow o} = \lambda_o \dfrac{T_i}{T_o} \dfrac{G(1)}{G\left(\sqrt{T_i/T_o}\right)},
        \end{equation}
        where $G(x)=(\text{erf}(x)-x\cdot\text{erf}^{\prime}(x))/2x^2$, and erf is the error function.
        We found $\lambda_{i\rightarrow o}=2.84$\,kpc.
        The same estimation applied to the other two wedges results in $\lambda_{i\rightarrow o} = 2.34$\,kpc for the SW, and $\lambda_{i\rightarrow o} = 1.66$\,kpc for the S.

        To determine the width of the edge, we fit a Gaussian broadened broken (or double broken) power-law to the SB profiles in all three directions. We assume the two edges in W have the same width.  
        The widths are estimated to be $\sigma =23.4^{+20.1}_{-15.0}$\,kpc, $3.3^{+4.2}_{-2.8}$\,kpc and $ 11.7^{+5.6}_{-5.6}$\,kpc for the S, SW, and W, respectively.
        The width of the SW is only $1.4$ times the $\lambda_{i\rightarrow o}$, the sharpest of all three, indicating a suppressed diffusion along this direction.
        The widths in the other two directions are unconstrained due to degraded fit quality when including Gaussian broadening.
        Altogether, we observe a clear azimuthal variation across the three directions with the sharpest edge in the SW, and disrupted edges toward the other two directions along the cold front.

     \subsection{Surface Brightness Fluctuations}

\begin{figure*}
            \centering
            \includegraphics[width=\textwidth]{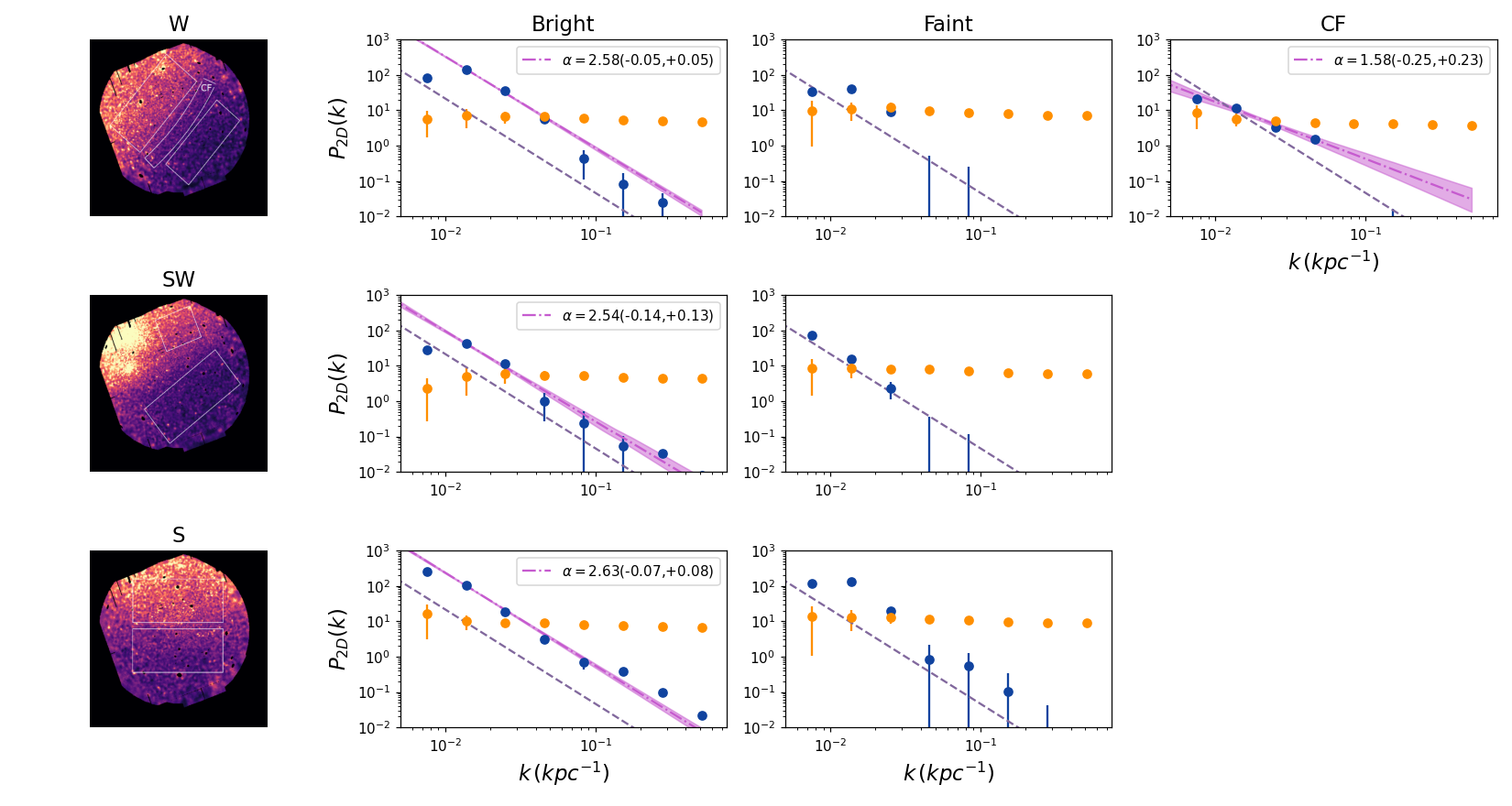}
            \caption{Resulting power spectra extracted from the bright and faint sides of the cold front.
            The W wedge has one more sub-region, CF, corresponding to the region between the detected double edges.
            The power spectra are shown in blue, which are corrected for various biases, including Poisson noise, normalization bias, and PSF bias.
            The Poisson noise is shown in orange points distributed in a flat power spectrum.
            The best-fit power-law model is shown in the purple solid line with the shaded area indicating the $1\sigma$ errors, estimated using the MCMC method.
            The dashed line shows the 2D Kolmogorov spectrum (slope $\alpha=8/3$) for comparison.
            We show the best-fit results only if the fitting converges.
            The upper bound of the x-axis marks the XMM resolution (half energy width HEW $\sim 15\,''$).
            }
            \label{fig:psdresults301}
        \end{figure*}

        A feasible way to probe KHI-induced turbulence is through the power spectra of the surface brightness fluctuations \citep{2012MNRAS.421.1123C}.
        To extract the power spectrum from an image, we adopted the modified $\Delta$-variance method described in \cite{2012MNRAS.426.1793A}.
        In brief, a Mexican hat filter was applied to images to screen for the power of the surface brightness fluctuations at a specific pixel scale $\sigma$, and suppress the powers at other scales.
        The power spectra of fluctuations at a given wavenumber $k=1/(\sqrt{2 \pi^2} \sigma)$ are then obtained from the square of the convolved images $I_{\sigma}$ multiplied by a few scaling factors,
        \begin{equation}
            P(k) = \dfrac{1}{\pi \epsilon^2 k^2 \mu_{s}^2} \dfrac{\sum I_{\sigma}^2}{\sum E^2},
        \end{equation}
        where $E$ is the exposure map and $\mu_s$ is the mean surface brightness of the region.
        Following \cite{2012MNRAS.426.1793A}, the dimensionless factor $\epsilon$, representing the small difference between the widths of two Gaussians in the Mexican hat function, is fixed at $0.001$.
        This analysis was applied to the bright and faint sides of the cold front in the three regions, as shown in the first column of images in Fig.\,\ref{fig:psdresults301}.
        An additional sub-region, referred to as CF, in the W wedge was used to extract the power spectrum of the region between the two SB edges. 
    
        The extracted power spectra are the combined product of signal and noise spectra.  
        One source of systematic uncertainties comes from the Poisson noise of astronomical X-ray images \citep{2012MNRAS.421.1123C}.
        In principle, the Poisson noise has no spectral features, and thus can be approximated as a flat power spectrum,
        \begin{equation}
            P_{poisson} = \dfrac{\sum I}{\mu_s^2 \sum E^2}.
        \end{equation}
        Numerically, the Poisson noise in an image is estimated using the underlying surface brightness of the cluster emission together with the background noise \citep{2017A&A...605A..25E}.
        To acquire robust estimates of the noise, we measured the corresponding uncertainty via a set of $100$ Markov Chain Monte Carlo (MCMC) realizations from the surface brightness profile of the Virgo Cluster.
        The MCMC algorithm was implemented from \textbf{emcee} package \citep{2013PASP..125..306F}.
        The resulting Poisson spectrum was directly subtracted from the ``raw" power spectrum.
        Note that the uncertainty of the signal power spectra has the same magnitude as that of the Poisson noise \citep{2023ApJ...951...41R}.
        Finally, we corrected for the bias introduced by the point spread function (PSF) of \xmm as well as the deviation between the power spectrum and that recovered by the modified $\Delta$-variance method, as described in Appendix\,\ref{sec:psf}.

        The slope of the power spectrum is an important diagnostic of the turbulent cascade \citep{2013A&A...559A..78G}.
        We fitted the power spectrum with a powerlaw,
        \begin{equation}
            F(k) = P_0 k^{-\alpha}.
        \end{equation}
        The best-fit values of the parameters were found using a non-linear regression algorithm.
        The range was set to $k<0.1\,\text{kpc}^{-1}$, which roughly corresponds to the scales above the Poisson-dominant regime.
        The uncertainties of model parameters were estimated through \textbf{emcee}.
        In Fig.\,\ref{fig:psdresults301}, we show the power spectra together with the Poisson spectra, and the best-fit models. For the bright side, we find slopes of power spectra consistent with a 2D Kolmogorov spectrum.
        The slopes on the faint side are not constrained due to a low signal-to-noise ratio.
        The power spectrum within the double edges of the W wedge shows a shallower slope, similar to the tails of infalling substructures into clusters, where flattening has been attributed to suppression of effective conduction \citep{2017A&A...605A..25E, 2023MNRAS.522.2105M}.

\section{Summary}
\label{sec:conclude}

    In this work, we presented \xmm observations of an ancient sloshing cold front at $r\approx 250$\,kpc (0.25\,$R_{\rm vir}$) in the Virgo Cluster, which fills the gap between the two well-studied cold fronts at the center of Virgo and an ancient sloshing cold front at half virial radius of Perseus. The analysis was complemented by a numerical simulation of a binary cluster merger. We summarize the main findings of this work as follows:

    \begin{enumerate}
        \item We compare the position and direction of this outer cold front and those of the two inner cold fronts with the simulation, which suggests that all three sloshing cold fronts may have been created by the off-axis merger of M49. This outer cold front is the oldest, likely created $2$\,--\,$3$ Gyr ago.
        \item The surface brightness profiles show that the south and southwest wedges of this cold front have a single surface brightness jump, while the west wedge exhibits double edges. The transition from single to double edges may imply the presence of a split in the cold front, which may indicate the presence of KHI and favor the suppression of viscosity in the ICM.
        \item The \textit{deprojected} thermal profiles are consistent with expectations for sloshing cold fronts: density decreases and temperature increases across the surface brightness edge, while pressure remains approximately continuous across the front. However, at radii larger than the fainter and hotter side of the cold front, we observe a pressure drop,
        steeper than the global pressure profile.
        \color{black}
        The pressure difference may indicate additional non-thermal pressure support at cluster outskirts.
        \item The intrinsic width of the SW portion (leading edge) of the cold front is only $1.4$ times the mean free path of the ICM, suggesting a suppression of transport processes, likely by the aligned magnetic field amplified by the sloshing.
        \item The Fe abundance of both the bright and faint sides are consistent with 0.3\,Z$_{\odot}$. The abundance ratios on both sides appear to be sub Solar. Considering that the ICM on the bright side was brought to large radii from the cluster center, a lack of metallicity variation suggests a well-mixed ICM chemical distribution. 
        \item The power spectra of the surface brightness fluctuations of the bright side of the cold front are consistent with the 2D Kolmogorov spectrum, while those of the faint side are not constrained.
    \end{enumerate}

Moving forward, mapping out the full structure of ancient sloshing cold fronts will be crucial for understanding their evolution and the merger histories of galaxy clusters.
Next-generation X-ray observatories with wide field of view, such as the proposed Advanced X-ray Imaging Satellite \citep[AXIS;][]{2023SPIE12678E..1ER}, will advance the study of large-radius sloshing cold fronts, offering critical insights into the microphysical conditions of the ICM.


\section*{Acknowledgment}
We acknowledge support from the Smithsonian Institution, the Chandra High Resolution Camera Project through NASA contract NAS8-03060, and NASA Grants 80NSSC19K0116, 80NSSC24K1776 and GO1-22132X.

\appendix
\section{\xmm background modeling}
\label{sec:bkg}
 The astrophysical background (AXB) consists of the galactic halo (GH) emission, the local hot bubble (LHB), and the unresolved point sources (CXB).
        The GH emission was modeled as an absorbed plasma model \texttt{apec}$_{\text{GH}}$ with temperature fixed at $0.2$\,keV, redshift fixed at $z=0$, and a metallicity of $Z=1$\,$Z_{\odot}$.
        The LHB was modeled as an $0.104$\,keV unabsorbed plasma \texttt{apec}$_{\text{LHB}}$ with the same redshift and metallicity as \texttt{apec}$_{\text{GH}}$.
        The unresovled point sources were characterized by the absorbed power-law \texttt{powerlaw}$_{\text{CXB}}$ with the photon index of $1.5$. 
        The model parameters listed in Table~\ref{tab:axb} were adopted from \cite{2017MNRAS.469.1476S} and fixed in the following analysis. We tested the effect of varying the CXB component as presented in Appendix\,\ref{sec:sys}.
        
        \begin{table}
            \centering
            \caption{Astrophysical X-ray background model parameters. The values are taken from \cite{2017MNRAS.469.1476S}}
            \begin{tabular}{ccc}
                Component & kT (keV)/$\Gamma$ & norm $\times 10^{-6}$ (W/S)\\
                \hline
                CXB & 1.5 & 0.89 \\
                GH & 0.2 & 1.27/2.33 \\
                LHB & 0.104 & 1.35/1.46 \\
            \end{tabular}
            \label{tab:axb}
        \end{table}

        The NXB can be approximated by a combination of a set of Gaussians \texttt{gauss} representing the fluorescent lines plus a broken power-law model \texttt{bknpower} for the continuum component with the energy break fixed at $3$\,keV.
        For \texttt{gauss} parts, the line energies were taken from \cite{2017ApJ...851...69S}, and the widths only varied within the range of $0.3$\,keV.
        For \texttt{bknpower}, we only restrict the break energy at $3$\,keV, and set other parameters free.
        
        The level of remaining soft proton flares after filtering was estimated by comparing the in-field and unexposured corner count rates in the energy of $6$--$12$\,keV.
        If the ratio F$_\text{in}$/F$_\text{out} > 1.5$, we included an additional power-law component to take into account the residual of the soft proton.

        We utilize the FWC spectra in the fitting to constrain the aforementioned NXB components.
        The FWC files were chosen with observation dates matching closest to our observations.
        Following \cite{2017ApJ...851...69S}, both the FWC and the observation spectra were simultaneously fit with a model containing AXB, NXB, and ICM. The NXB component was not folded through the ARFs. The AXB and ICM flux were fixed at 0 for the FWC spectra. The normalizations of the NXB components were linked between the FWC and observation spectra through flux ratios allowed to vary within $5\%$, with the mean values determined from the unexposed corners of the detectors.

\section{Possible systematics in spectral analysis}
\label{sec:sys}

        In addition to statistical uncertainties, there are systematics involved in our spectral analysis, including the variance of CXB flux and the discrepancy between MOS and pn detectors.
        To properly propagate the uncertainties of the CXB flux, we refitted all spectra with the normalization of \texttt{powerlaw}$_{\text{CXB}}$ varied by $\pm 20\%$.
        \begin{equation}
            \sigma_{\text{CXB}} = Y_{under} - Y_{over},
        \end{equation}
        where $Y_{i}$ represents a given parameter of \texttt{apec} or \texttt{vapec}.
        We found that varying CXB normalization by $20\%$ introduces $\approx 5\%$ systematic into the temperature, metallicity or normalization measurements.

        To address the discrepancy between MOS and pn spectra, we fitted the MOS (MOS1 and MOS2 jointly) and pn spectra separately.
        The discrepancies were calculated by the error-subtracted variance \citep[or excess variance,][]{2015A&A...575A..37M},
        \begin{equation}
            \sigma_{det}^2 = \left< \Tilde{Y}^2 \right> - \left< \sigma_{err}^2 \right>,
        \end{equation}
        where $\Tilde{Y}=Y_{i} - \left<Y\right>$, and $\sigma_{err}$ is the statistical uncertainty estimated from individual detectors denoted by an index \textit{i}.
        Then, the first term on the right hand side is the variance of estimate calculated from individual detectors against the arithmetic mean of $\left< Y \right> = \left( Y_{\text{MOS}} + Y_{\text{pn}} \right) / 2$.
        Note that a negative error-subtracted variance is forced to be zero as the variance of estimate is insignificant compared to the statistical error terms.
        Finally, the total uncertainty can be combined quadratically,
        \begin{equation}
            \sigma_{tot} = \sqrt{\sigma_{stat}^2 + \sigma_{\text{CXB}}^2 + \sigma_{det}^2},
        \end{equation}
        where $\sigma_{stat}$ in the total uncertainty is the statistical uncertainty from the joint fitting of all three instruments.
 
\section{Biases in power spectrum analysis and corrections}
\label{sec:psf}
        We correct the extracted spectra from the point spread function (PSF) bias.
        \cite{2023ApJ...951...41R} provided a numerical approximation to the XMM PSF using a triple Gaussian function to correct the bias from the power spectra directly.
        The Gaussian function is a convenient approximation due to its invariant property under the Fourier transform, and thus, provides a direct way to express the PSFs in Fourier space.
        We follow \cite{2023ApJ...951...41R} to estimate the power spectra of the PSFs.
        The PSF per position in detector space of a single observation can be generated by the task \texttt{psfgen} from the XMM-ESAS package.
        We select in each region for each observation three different positions, one at the region center and two on the sides, to generate the PSF images for all three detectors.
        The spectra of PSFs are then extracted in the same way as the observations, and fitted directly with a triple Gaussian function.
        The triple Gaussian function has the form,
        \begin{equation}
            G(k) = \sum_{i=1}^3 c_{i} e^{-k^2 / k_i^2},
        \end{equation}
        where $c_i$ is the normalization of each Gaussian such that $\sum_{i} c_i = 1$, and $k_i$ is the width of a Gaussian.
        As a result, the correction term for the PSF bias can be expressed by,
        \begin{equation}
            \dfrac{P_{corr}}{P_{unc}}(k) = \sum_{i=1}^{3}\sum_{j=1}^{3} c_i c_j \left( \dfrac{2 x_i^2 x_j^2 + x_i^2 + x_j^2}{2 x_i^2 x_j^2}\right)^{n/2+2-\alpha/2},
        \end{equation}
        where $P_{corr}(k)$ and $P_{unc}(k)$ are the corrected and uncorrected spectra, respectively.
        The term $x_i$ is defined as $x_i = k_i / k$, where $k$ is the Fourier mode of spectrum.
        In each region of each observation, we take a median value of these correction terms as they do not vary much across detectors as well as from the region center to the edges.
        
        Another critical systematic emerges from the deviation between a powerlaw spectrum $P(k)$ and the spectrum $P_{rec}(k)$ recovered by the modified $\Delta$-variance method.
        Following the prescription by \cite{2012MNRAS.426.1793A}, we corrected the spectra from this normalization bias by,
        \begin{equation}
            \dfrac{P_{rec}(k)}{P(k)} = 2^{\alpha/2} \dfrac{\Gamma\left(n/2+2-\alpha/2\right)}{\Gamma\left(n/2+2\right)},
        \end{equation}
        where $n$ is the dimension of data, and $\alpha$ is the slope of the powerlaw.

\bibliography{main}{}

\begin{thebibliography}{}
\expandafter\ifx\csname natexlab\endcsname\relax\def\natexlab#1{#1}\fi
\providecommand{\url}[1]{\href{#1}{#1}}
\providecommand{\dodoi}[1]{doi:~\href{http://doi.org/#1}{\nolinkurl{#1}}}
\providecommand{\doeprint}[1]{\href{http://ascl.net/#1}{\nolinkurl{http://ascl.net/#1}}}
\providecommand{\doarXiv}[1]{\href{https://arxiv.org/abs/#1}{\nolinkurl{https://arxiv.org/abs/#1}}}

\bibitem[{{Allen} {et~al.}(2004){Allen}, {Schmidt}, {Ebeling}, {Fabian}, \& {van Speybroeck}}]{2004MNRAS.353..457A}
{Allen}, S.~W., {Schmidt}, R.~W., {Ebeling}, H., {Fabian}, A.~C., \& {van Speybroeck}, L. 2004, \mnras, 353, 457, \dodoi{10.1111/j.1365-2966.2004.08080.x}

\bibitem[{{Ar{\'e}valo} {et~al.}(2012){Ar{\'e}valo}, {Churazov}, {Zhuravleva}, {Hern{\'a}ndez-Monteagudo}, \& {Revnivtsev}}]{2012MNRAS.426.1793A}
{Ar{\'e}valo}, P., {Churazov}, E., {Zhuravleva}, I., {Hern{\'a}ndez-Monteagudo}, C., \& {Revnivtsev}, M. 2012, \mnras, 426, 1793, \dodoi{10.1111/j.1365-2966.2012.21789.x}

\bibitem[{{Arnaud}(1996)}]{1996ASPC..101...17A}
{Arnaud}, K.~A. 1996, in Astronomical Society of the Pacific Conference Series, Vol. 101, Astronomical Data Analysis Software and Systems V, ed. G.~H. {Jacoby} \& J.~{Barnes}, 17

\bibitem[{{CHEX-MATE Collaboration} {et~al.}(2021){CHEX-MATE Collaboration}, {Arnaud}, {Ettori}, {Pratt}, {Rossetti}, {Eckert}, {Gastaldello}, {Gavazzi}, {Kay}, {Lovisari}, {Maughan}, {Pointecouteau}, {Sereno}, {Bartalucci}, {Bonafede}, {Bourdin}, {Cassano}, {Duffy}, {Iqbal}, {Maurogordato}, {Rasia}, {Sayers}, {Andrade-Santos}, {Aussel}, {Barnes}, {Barrena}, {Borgani}, {Burkutean}, {Clerc}, {Corasaniti}, {Cuillandre}, {De Grandi}, {De Petris}, {Dolag}, {Donahue}, {Ferragamo}, {Gaspari}, {Ghizzardi}, {Gitti}, {Haines}, {Jauzac}, {Johnston-Hollitt}, {Jones}, {K{\'e}ruzor{\'e}}, {Le Brun}, {Mayet}, {Mazzotta}, {Melin}, {Molendi}, {Nonino}, {Okabe}, {Paltani}, {Perotto}, {Pires}, {Radovich}, {Rubino-Martin}, {Salvati}, {Saro}, {Sartoris}, {Schellenberger}, {Streblyanska}, {Tarr{\'\i}o}, {Tozzi}, {Umetsu}, {van der Burg}, {Vazza}, {Venturi}, {Yepes}, \& {Zarattini}}]{2021A&A...650A.104C}
{CHEX-MATE Collaboration}, {Arnaud}, M., {Ettori}, S., {et~al.} 2021, \aap, 650, A104, \dodoi{10.1051/0004-6361/202039632}

\bibitem[{{Churazov} {et~al.}(2003){Churazov}, {Forman}, {Jones}, \& {B{\"o}hringer}}]{2003ApJ...590..225C}
{Churazov}, E., {Forman}, W., {Jones}, C., \& {B{\"o}hringer}, H. 2003, \apj, 590, 225, \dodoi{10.1086/374923}

\bibitem[{{Churazov} {et~al.}(2012){Churazov}, {Vikhlinin}, {Zhuravleva}, {Schekochihin}, {Parrish}, {Sunyaev}, {Forman}, {B{\"o}hringer}, \& {Randall}}]{2012MNRAS.421.1123C}
{Churazov}, E., {Vikhlinin}, A., {Zhuravleva}, I., {et~al.} 2012, \mnras, 421, 1123, \dodoi{10.1111/j.1365-2966.2011.20372.x}

\bibitem[{Dembinski {et~al.}(2020)Dembinski, Ongmongkolkul, {et~al.}}]{iminuit}
Dembinski, H., Ongmongkolkul, P., {et~al.} 2020, \dodoi{10.5281/zenodo.3949207}

\bibitem[{{Dorman} \& {Arnaud}(2001)}]{2001ASPC..238..415D}
{Dorman}, B., \& {Arnaud}, K.~A. 2001, in Astronomical Society of the Pacific Conference Series, Vol. 238, Astronomical Data Analysis Software and Systems X, ed. J.~{Harnden}, F.~R., F.~A. {Primini}, \& H.~E. {Payne}, 415

\bibitem[{{Eckert} {et~al.}(2020){Eckert}, {Finoguenov}, {Ghirardini}, {Grandis}, {Kaefer}, {Sanders}, \& {Ramos-Ceja}}]{2020OJAp....3E..12E}
{Eckert}, D., {Finoguenov}, A., {Ghirardini}, V., {et~al.} 2020, The Open Journal of Astrophysics, 3, 12, \dodoi{10.21105/astro.2009.13944}

\bibitem[{{Eckert} {et~al.}(2017){Eckert}, {Gaspari}, {Owers}, {Roediger}, {Molendi}, {Gastaldello}, {Paltani}, {Ettori}, {Venturi}, {Rossetti}, \& {Rudnick}}]{2017A&A...605A..25E}
{Eckert}, D., {Gaspari}, M., {Owers}, M.~S., {et~al.} 2017, \aap, 605, A25, \dodoi{10.1051/0004-6361/201730555}

\bibitem[{{Foreman-Mackey} {et~al.}(2013){Foreman-Mackey}, {Hogg}, {Lang}, \& {Goodman}}]{2013PASP..125..306F}
{Foreman-Mackey}, D., {Hogg}, D.~W., {Lang}, D., \& {Goodman}, J. 2013, \pasp, 125, 306, \dodoi{10.1086/670067}

\bibitem[{{Gaspari} \& {Churazov}(2013)}]{2013A&A...559A..78G}
{Gaspari}, M., \& {Churazov}, E. 2013, \aap, 559, A78, \dodoi{10.1051/0004-6361/201322295}

\bibitem[{{Ichinohe} {et~al.}(2017){Ichinohe}, {Simionescu}, {Werner}, \& {Takahashi}}]{2017MNRAS.467.3662I}
{Ichinohe}, Y., {Simionescu}, A., {Werner}, N., \& {Takahashi}, T. 2017, \mnras, 467, 3662, \dodoi{10.1093/mnras/stx280}

\bibitem[{{Kalberla} {et~al.}(2005){Kalberla}, {Burton}, {Hartmann}, {Arnal}, {Bajaja}, {Morras}, \& {P{\"o}ppel}}]{2005A&A...440..775K}
{Kalberla}, P.~M.~W., {Burton}, W.~B., {Hartmann}, D., {et~al.} 2005, \aap, 440, 775, \dodoi{10.1051/0004-6361:20041864}

\bibitem[{{Lin} {et~al.}(2024){Lin}, {Su}, {Gastaldello}, \& {Jacobs}}]{2024ApJ...977..176L}
{Lin}, S.-C., {Su}, Y., {Gastaldello}, F., \& {Jacobs}, N. 2024, \apj, 977, 176, \dodoi{10.3847/1538-4357/ad8888}

\bibitem[{{Lin} {et~al.}(2022){Lin}, {Su}, {Liang}, {Zhang}, {Jacobs}, \& {Zhang}}]{2022MNRAS.512.3885L}
{Lin}, S.-C., {Su}, Y., {Liang}, G., {et~al.} 2022, \mnras, 512, 3885, \dodoi{10.1093/mnras/stac725}

\bibitem[{{Lodders}(2003)}]{2003ApJ...591.1220L}
{Lodders}, K. 2003, \apj, 591, 1220, \dodoi{10.1086/375492}

\bibitem[{{Lyutikov}(2006)}]{2006MNRAS.373...73L}
{Lyutikov}, M. 2006, \mnras, 373, 73, \dodoi{10.1111/j.1365-2966.2006.10835.x}

\bibitem[{{Markevitch} \& {Vikhlinin}(2007)}]{2007PhR...443....1M}
{Markevitch}, M., \& {Vikhlinin}, A. 2007, \physrep, 443, 1, \dodoi{10.1016/j.physrep.2007.01.001}

\bibitem[{{McCall} {et~al.}(2024){McCall}, {Reiprich}, {Veronica}, {Pacaud}, {Sanders}, {Edler}, {Br{\"u}ggen}, {Bulbul}, {de Gasperin}, {Gatuzz}, {Liu}, {Merloni}, {Migkas}, \& {Zhang}}]{2024A&A...689A.113M}
{McCall}, H., {Reiprich}, T.~H., {Veronica}, A., {et~al.} 2024, \aap, 689, A113, \dodoi{10.1051/0004-6361/202449391}

\bibitem[{{Mei} {et~al.}(2007){Mei}, {Blakeslee}, {C{\^o}t{\'e}}, {Tonry}, {West}, {Ferrarese}, {Jord{\'a}n}, {Peng}, {Anthony}, \& {Merritt}}]{2007ApJ...655..144M}
{Mei}, S., {Blakeslee}, J.~P., {C{\^o}t{\'e}}, P., {et~al.} 2007, \apj, 655, 144, \dodoi{10.1086/509598}

\bibitem[{{Mernier} {et~al.}(2015){Mernier}, {de Plaa}, {Lovisari}, {Pinto}, {Zhang}, {Kaastra}, {Werner}, \& {Simionescu}}]{2015A&A...575A..37M}
{Mernier}, F., {de Plaa}, J., {Lovisari}, L., {et~al.} 2015, \aap, 575, A37, \dodoi{10.1051/0004-6361/201425282}

\bibitem[{{Mirakhor} {et~al.}(2023){Mirakhor}, {Walker}, \& {Runge}}]{2023MNRAS.522.2105M}
{Mirakhor}, M.~S., {Walker}, S.~A., \& {Runge}, J. 2023, \mnras, 522, 2105, \dodoi{10.1093/mnras/stad1088}

\bibitem[{{Owers} {et~al.}(2009){Owers}, {Nulsen}, {Couch}, \& {Markevitch}}]{2009ApJ...704.1349O}
{Owers}, M.~S., {Nulsen}, P. E.~J., {Couch}, W.~J., \& {Markevitch}, M. 2009, \apj, 704, 1349, \dodoi{10.1088/0004-637X/704/2/1349}

\bibitem[{{Pratt} {et~al.}(2010){Pratt}, {Arnaud}, {Piffaretti}, {B{\"o}hringer}, {Ponman}, {Croston}, {Voit}, {Borgani}, \& {Bower}}]{2010A&A...511A..85P}
{Pratt}, G.~W., {Arnaud}, M., {Piffaretti}, R., {et~al.} 2010, \aap, 511, A85, \dodoi{10.1051/0004-6361/200913309}

\bibitem[{{Reynolds} {et~al.}(2023){Reynolds}, {Kara}, {Mushotzky}, {Ptak}, {Koss}, {Williams}, {Allen}, {Bauer}, {Bautz}, {Bogadhee}, {Burdge}, {Cappelluti}, {Cenko}, {Chartas}, {Chan}, {Corrales}, {Daylan}, {Falcone}, {Foord}, {Grant}, {Habouzit}, {Haggard}, {Herrmann}, {Hodges-Kluck}, {Kargaltsev}, {King}, {Kounkel}, {Lopez}, {Marchesi}, {McDonald}, {Meyer}, {Miller}, {Nynka}, {Okajima}, {Pacucci}, {Russell}, {Safi-Harb}, {Strassun}, {Trindade Falc{\~a}o}, {Walker}, {Wilms}, {Yukita}, \& {Zhang}}]{2023SPIE12678E..1ER}
{Reynolds}, C.~S., {Kara}, E.~A., {Mushotzky}, R.~F., {et~al.} 2023, in Society of Photo-Optical Instrumentation Engineers (SPIE) Conference Series, Vol. 12678, UV, X-Ray, and Gamma-Ray Space Instrumentation for Astronomy XXIII, ed. O.~H. {Siegmund} \& K.~{Hoadley}, 126781E, \dodoi{10.1117/12.2677468}

\bibitem[{{Roediger} {et~al.}(2013){Roediger}, {Kraft}, {Forman}, {Nulsen}, \& {Churazov}}]{2013ApJ...764...60R}
{Roediger}, E., {Kraft}, R.~P., {Forman}, W.~R., {Nulsen}, P.~E.~J., \& {Churazov}, E. 2013, \apj, 764, 60, \dodoi{10.1088/0004-637X/764/1/60}

\bibitem[{{Romero} {et~al.}(2023){Romero}, {Gaspari}, {Schellenberger}, {Bhandarkar}, {Devlin}, {Dicker}, {Forman}, {Khatri}, {Kraft}, {Di Mascolo}, {Mason}, {Moravec}, {Mroczkowski}, {Nulsen}, {Orlowski-Scherer}, {Perez Sarmiento}, {Sarazin}, {Sievers}, \& {Su}}]{2023ApJ...951...41R}
{Romero}, C.~E., {Gaspari}, M., {Schellenberger}, G., {et~al.} 2023, \apj, 951, 41, \dodoi{10.3847/1538-4357/acd3f0}

\bibitem[{{Sanders} \& {Fabian}(2007)}]{2007MNRAS.381.1381S}
{Sanders}, J.~S., \& {Fabian}, A.~C. 2007, \mnras, 381, 1381, \dodoi{10.1111/j.1365-2966.2007.12347.x}

\bibitem[{{Sanders} {et~al.}(2016){Sanders}, {Fabian}, {Taylor}, {Russell}, {Blundell}, {Canning}, {Hlavacek-Larrondo}, {Walker}, \& {Grimes}}]{2016MNRAS.457...82S}
{Sanders}, J.~S., {Fabian}, A.~C., {Taylor}, G.~B., {et~al.} 2016, \mnras, 457, 82, \dodoi{10.1093/mnras/stv2972}

\bibitem[{{Sarkar} {et~al.}(2022){Sarkar}, {Su}, {Truong}, {Randall}, {Mernier}, {Gastaldello}, {Biffi}, \& {Kraft}}]{2022MNRAS.516.3068S}
{Sarkar}, A., {Su}, Y., {Truong}, N., {et~al.} 2022, \mnras, 516, 3068, \dodoi{10.1093/mnras/stac2416}

\bibitem[{{Schwarz}(1978)}]{1978AnSta...6..461S}
{Schwarz}, G. 1978, Annals of Statistics, 6, 461

\bibitem[{{Simionescu} {et~al.}(2010){Simionescu}, {Werner}, {Forman}, {Miller}, {Takei}, {B{\"o}hringer}, {Churazov}, \& {Nulsen}}]{2010MNRAS.405...91S}
{Simionescu}, A., {Werner}, N., {Forman}, W.~R., {et~al.} 2010, \mnras, 405, 91, \dodoi{10.1111/j.1365-2966.2010.16450.x}

\bibitem[{{Simionescu} {et~al.}(2017){Simionescu}, {Werner}, {Mantz}, {Allen}, \& {Urban}}]{2017MNRAS.469.1476S}
{Simionescu}, A., {Werner}, N., {Mantz}, A., {Allen}, S.~W., \& {Urban}, O. 2017, \mnras, 469, 1476, \dodoi{10.1093/mnras/stx919}

\bibitem[{{Simionescu} {et~al.}(2012){Simionescu}, {Werner}, {Urban}, {Allen}, {Fabian}, {Sanders}, {Mantz}, {Nulsen}, \& {Takei}}]{2012ApJ...757..182S}
{Simionescu}, A., {Werner}, N., {Urban}, O., {et~al.} 2012, \apj, 757, 182, \dodoi{10.1088/0004-637X/757/2/182}

\bibitem[{{Snowden} {et~al.}(2004){Snowden}, {Collier}, \& {Kuntz}}]{2004ApJ...610.1182S}
{Snowden}, S.~L., {Collier}, M.~R., \& {Kuntz}, K.~D. 2004, \apj, 610, 1182, \dodoi{10.1086/421841}

\bibitem[{{Spasic} {et~al.}(2024){Spasic}, {Edler}, {Su}, {Br{\"u}ggen}, {de Gasperin}, {Pasini}, {Heesen}, {Simonte}, {Boselli}, {R{\"o}ttgering}, \& {Fossati}}]{2024A&A...690A.195S}
{Spasic}, A., {Edler}, H.~W., {Su}, Y., {et~al.} 2024, \aap, 690, A195, \dodoi{10.1051/0004-6361/202450365}

\bibitem[{{Su} {et~al.}(2023){Su}, {Combes}, {Olivares}, {Castignani}, {Torne}, \& {van Weeren}}]{2023MNRAS.526.6052S}
{Su}, Y., {Combes}, F., {Olivares}, V., {et~al.} 2023, \mnras, 526, 6052, \dodoi{10.1093/mnras/stad3172}

\bibitem[{{Su} {et~al.}(2017{\natexlab{a}}){Su}, {Nulsen}, {Kraft}, {Roediger}, {ZuHone}, {Jones}, {Forman}, {Sheardown}, {Irwin}, \& {Randall}}]{2017ApJ...851...69S}
{Su}, Y., {Nulsen}, P. E.~J., {Kraft}, R.~P., {et~al.} 2017{\natexlab{a}}, \apj, 851, 69, \dodoi{10.3847/1538-4357/aa989e}

\bibitem[{{Su} {et~al.}(2017{\natexlab{b}}){Su}, {Kraft}, {Roediger}, {Nulsen}, {Forman}, {Churazov}, {Randall}, {Jones}, \& {Machacek}}]{2017ApJ...834...74S}
{Su}, Y., {Kraft}, R.~P., {Roediger}, E., {et~al.} 2017{\natexlab{b}}, \apj, 834, 74, \dodoi{10.3847/1538-4357/834/1/74}

\bibitem[{{Su} {et~al.}(2019){Su}, {Kraft}, {Nulsen}, {Jones}, {Maccarone}, {Mernier}, {Lovisari}, {Sheardown}, {Randall}, {Roediger}, {Fish}, {Forman}, \& {Churazov}}]{2019AJ....158....6S}
{Su}, Y., {Kraft}, R.~P., {Nulsen}, P.~E.~J., {et~al.} 2019, \aj, 158, 6, \dodoi{10.3847/1538-3881/ab1d51}

\bibitem[{{Tonry} {et~al.}(2001){Tonry}, {Dressler}, {Blakeslee}, {Ajhar}, {Fletcher}, {Luppino}, {Metzger}, \& {Moore}}]{2001ApJ...546..681T}
{Tonry}, J.~L., {Dressler}, A., {Blakeslee}, J.~P., {et~al.} 2001, \apj, 546, 681, \dodoi{10.1086/318301}

\bibitem[{{Urban} {et~al.}(2011){Urban}, {Werner}, {Simionescu}, {Allen}, \& {B{\"o}hringer}}]{2011MNRAS.414.2101U}
{Urban}, O., {Werner}, N., {Simionescu}, A., {Allen}, S.~W., \& {B{\"o}hringer}, H. 2011, \mnras, 414, 2101, \dodoi{10.1111/j.1365-2966.2011.18526.x}

\bibitem[{{Vaezzadeh} {et~al.}(2022){Vaezzadeh}, {Roediger}, {Cashmore}, {Hunt}, {ZuHone}, {Forman}, {Jones}, {Kraft}, {Nulsen}, {Su}, \& {Churazov}}]{2022MNRAS.514..518V}
{Vaezzadeh}, I., {Roediger}, E., {Cashmore}, C., {et~al.} 2022, \mnras, 514, 518, \dodoi{10.1093/mnras/stac784}

\bibitem[{{Walker} {et~al.}(2017){Walker}, {Hlavacek-Larrondo}, {Gendron-Marsolais}, {Fabian}, {Intema}, {Sanders}, {Bamford}, \& {van Weeren}}]{2017MNRAS.468.2506W}
{Walker}, S.~A., {Hlavacek-Larrondo}, J., {Gendron-Marsolais}, M., {et~al.} 2017, \mnras, 468, 2506, \dodoi{10.1093/mnras/stx640}

\bibitem[{{Walker} {et~al.}(2018){Walker}, {ZuHone}, {Fabian}, \& {Sanders}}]{2018NatAs...2..292W}
{Walker}, S.~A., {ZuHone}, J., {Fabian}, A., \& {Sanders}, J. 2018, Nature Astronomy, 2, 292, \dodoi{10.1038/s41550-018-0401-8}

\bibitem[{{Wen} {et~al.}(2009){Wen}, {Han}, \& {Liu}}]{2009ApJS..183..197W}
{Wen}, Z.~L., {Han}, J.~L., \& {Liu}, F.~S. 2009, \apjs, 183, 197, \dodoi{10.1088/0067-0049/183/2/197}

\bibitem[{{Werner} {et~al.}(2016){Werner}, {ZuHone}, {Zhuravleva}, {Ichinohe}, {Simionescu}, {Allen}, {Markevitch}, {Fabian}, {Keshet}, {Roediger}, {Ruszkowski}, \& {Sanders}}]{2016MNRAS.455..846W}
{Werner}, N., {ZuHone}, J.~A., {Zhuravleva}, I., {et~al.} 2016, \mnras, 455, 846, \dodoi{10.1093/mnras/stv2358}

\bibitem[{{ZuHone} \& {Su}(2022)}]{2022hxga.book...93Z}
{ZuHone}, J., \& {Su}, Y. 2022, in Handbook of X-ray and Gamma-ray Astrophysics, 93, \dodoi{10.1007/978-981-16-4544-0_124-1}

\bibitem[{{ZuHone} {et~al.}(2010){ZuHone}, {Markevitch}, \& {Johnson}}]{2010ApJ...717..908Z}
{ZuHone}, J.~A., {Markevitch}, M., \& {Johnson}, R.~E. 2010, \apj, 717, 908, \dodoi{10.1088/0004-637X/717/2/908}

\bibitem[{{ZuHone} {et~al.}(2011){ZuHone}, {Markevitch}, \& {Lee}}]{2011ApJ...743...16Z}
{ZuHone}, J.~A., {Markevitch}, M., \& {Lee}, D. 2011, \apj, 743, 16, \dodoi{10.1088/0004-637X/743/1/16}

\bibitem[{{Zuhone} \& {Roediger}(2016)}]{2016JPlPh..82c5301Z}
{Zuhone}, J.~A., \& {Roediger}, E. 2016, Journal of Plasma Physics, 82, 535820301, \dodoi{10.1017/S0022377816000544}

\end{thebibliography}
\bibliographystyle{aasjournal}

\end{document}